# How player and opponent personalities influence cooperative gameplay

Konstantina Ntretska[1], Nikos Avrantinis[1,2], George Tsatiris[1,3], Kostas Karpouzis[4]

[1] *School of Game Programming, SAE Athens, Greece*
[2] *Knowledge Engineering Lab, University of Piraeus, Greece*
[3] *Artificial Intelligence and Learning Systems Laboratory, National Technical University of Athens, Greece*
[4] *Dept. of Communication, Media and Culture, Panteion University of Social and Political Science, Athens, Greece*

**Abstract**
Research has shown that digital game players often feel engagement and rapport with a game hero or character when they can channel their own ambitions and goals through the hero's journey in the game world; in essence, they feel a sense of accomplishment and fulfillment whenever they put the game mechanics to use to help the hero reach a positive ending to the game quests. In the case of cooperative gameplay, rapport also has to do with their perception of their peers' skills, gameplay style and behavior within the game. In this paper, we describe an experiment to identify whether matching players with different personalities, as characterized by the OCEAN or Big-5 personality model, can influence their player experience with a custom-made, cooperative game.

## 1. Introduction

Digital games constitute a powerful medium when it comes to evoking player expressivity and emotions. Thanks to the need for coordination, cooperation (in the case of multiplayer games) and lightning-fast decision making, often in the presence of uncertainty or lacking complete information about the game world, gameplay pushes players in terms of experiencing and (usually) expressing intense emotions. In many cases, the inhibition brought in by social play, mostly when players are physically located in the same room, is overcome by the immediacy of emotional expression. The content of that expression can be sometimes moderated by the characteristics of multi-party play: whether we know our teammate in real life, our feelings and predisposition towards them, and our perception of their performance and skill may influence the way we feel about them or how we express ourselves. The final expression is also affected by our personality and our mood at the time of gameplay; extrovert players may illustrate more active emotions, either positive or negative, while introverts may choose to internalize their disappointment towards their teammates, should they feel they underperform.

In our work, we developed a digital game called "Joint Effort to Escape" (JEFE) and used it to test different scenarios of cooperation between previously unaffiliated players. Our main research question involved whether the personality of each player, as estimated by a standardized test, along with the perception of skill, would influence their performance as a team, when playing the game and going through the tasks it involved. The following sections introduce the relevance of personality in game play and how researchers can receive estimates of different personality dimensions via questionnaires and present the game and the results we collected.

---

## 2. Personality theories for interactive media

One of the most dominant family of theories describing personality is that of *traits*. A trait can be defined as a (more or less) stable characteristic which results in a person behaving in a specific manner or showing specific responses to situations or stimuli. Allport, one of the first researchers to propose this concept, described traits as building blocks of personality, categorized in three classes: cardinal traits, those that persons organize their life around; central traits, i.e., those that provide the foundation of one's personality, and secondary traits, behavior patterns that seem to be expressed in specific situations and environments. Eysenck built on this approach to include dimensions of personality (extroversion and introversion) which also include how we approach and deal with or in the presence of other people. The difference of theories based on traits with other theories, such as those by Freud or Rogers, is that they offer means and instruments for measurement. The 'Five Factor' model belongs to the trait family of theories and may also be found in literature as OCEAN, after the five main personality traits, Openness, Conscientiousness, Extraversion, Agreeableness, and Neuroticism, with the last one often being replaced by Emotional Stability, its conceptual opposite. Research has shown that personality can often be correlated with how people prefer to consume digital content. In the case of video games, this has to do with a number of factors, such as choice of game genre, game aesthetics, and playing style. For instance, according to Chory and Goodboy, players with high neuroticism scores tend to choose more violent video game genres, such as first-person shooters, over role-playing games.

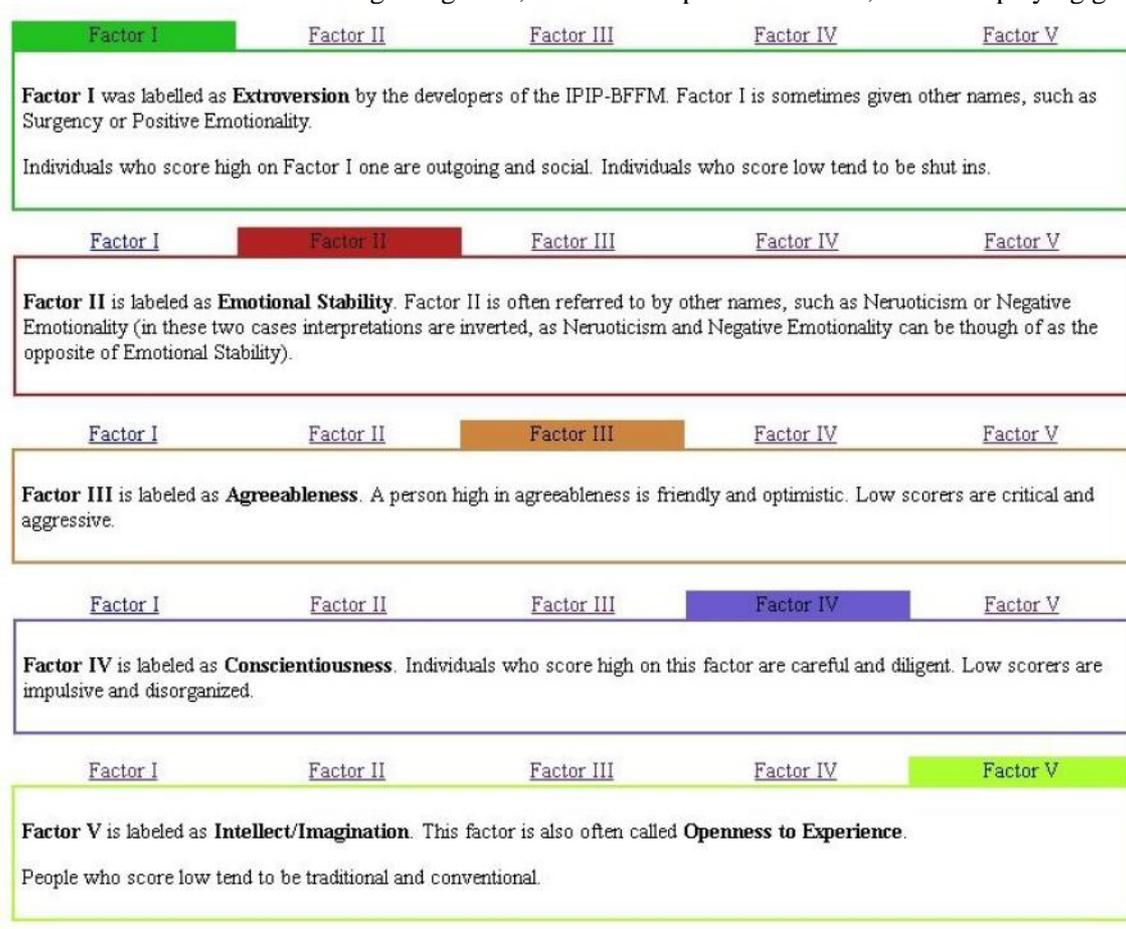

**Figure 1**: Description of the five factors in the OCEAN model, as shown in the questionnaire

In our work, personality measurement is an inherent concept of the research question. In most cases, this involves interviews conducted by experts and scoring the subject across different categories, corresponding to the use case and theoretical concepts at hand. Even though experts are professionals and have been taught how to consistently evaluate subjects, the outcome of this process is mostly subjective and carries on artefacts imposed by the presence of an evaluator in the same room as the subject (similar to the 'white coat' syndrome, which affects patients during medical evaluations, increasing their stress levels). Questionnaire-based evaluation alleviates this problem and, thanks to extensive calibration, appears to be more consistent and subjective than manual evaluation. In order to evaluate our players, we used Goldberg's 50-item International Personality Item Pool[2], which was completed prior to them playing the game.

## 3. The "Joint Effort to Escape" game

In our work, we focused on cooperative video games, or 'co-op'. These are games where two or more players participate in the same narrative and virtual world, and work together towards a common goal, as opposed to competitive games where every player or group of players play for themselves. This distinction runs across different genres, game mechanics and aesthetics, with puzzle games being the most popular co-op genre. In this framework, we developed a local (i.e., not network-based) co-op game and investigated whether player personalities influence player behavior and experience in terms of interaction, fun and engagement. Player personalities were also considered in terms of game play result, in the sense of identifying which combination of individual personality traits would result in solving the puzzles contained in the game.

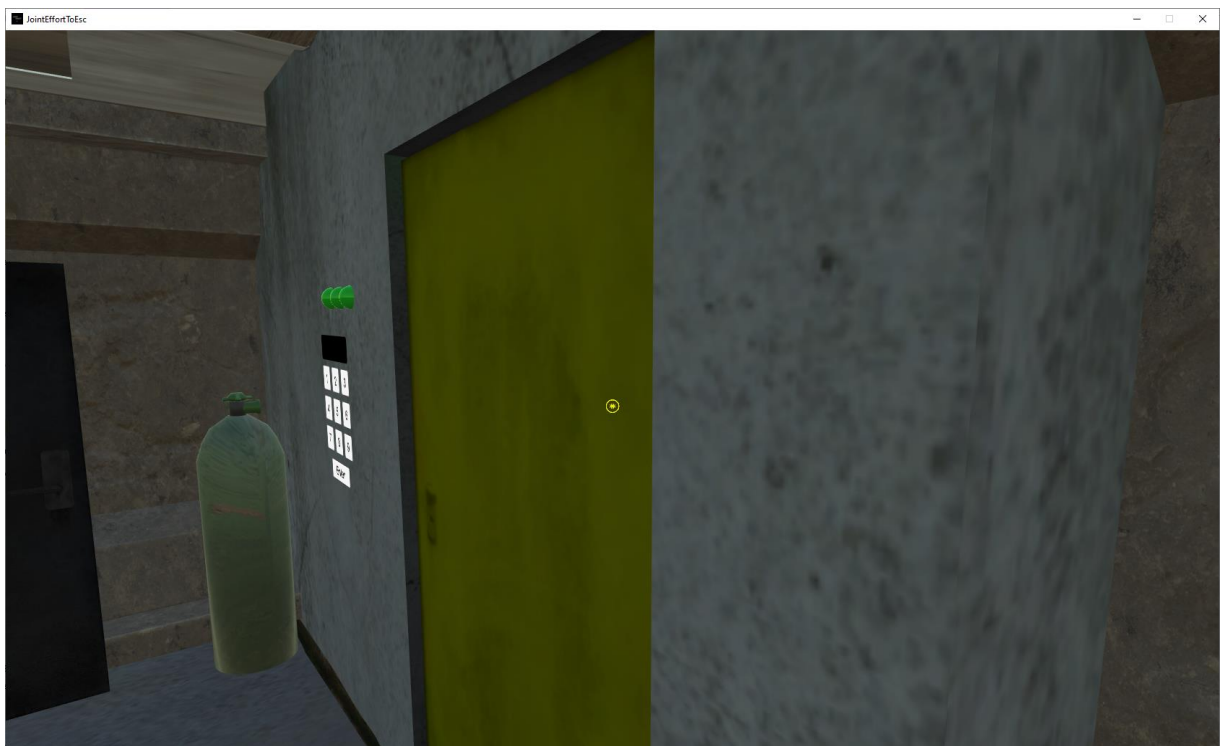

**Figure 2**: An escape scene in "Joint Effort to Escape"

All participants were millennials and characterized themselves as 'game players'; they also had no previous acquaintance, so as to eliminate (or, at least, reduce) any social predisposition effects. Before game play, participants were asked to complete the questionnaire described above, providing individual scores with respect to each of the five factors in the OCEAN model. Then, they played a cooperative game that we developed, called 'Joint Effort to Escape', where one of the players is in possession of all the information needed to escape a maze and must guide the other player to the exit.

---

[2] IPIP, A Scientific Collaboratory for the Development of Advanced Measures of Personality and Other Individual Differences, available at https://ipip.ori.org/

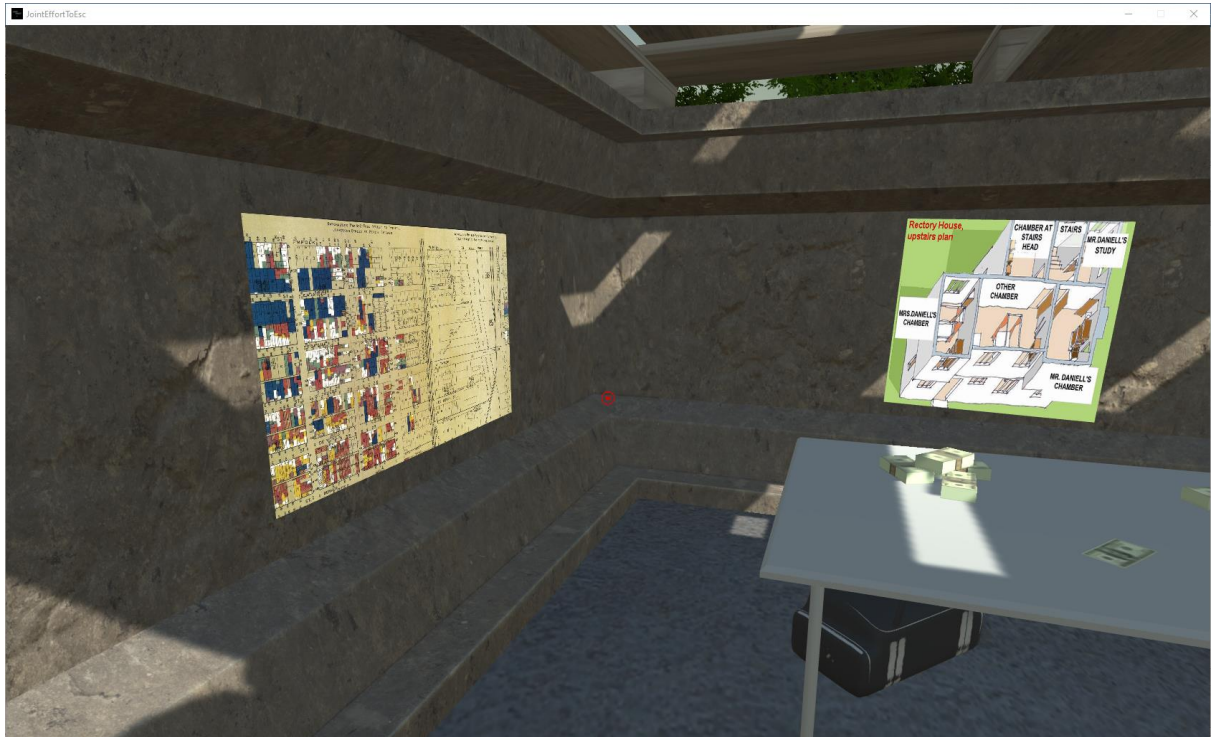

**Figure 3**: Information available to players to guide them through the maze.

## 4. Results and conclusions

Given that this was a cooperative game, players would win and lose as a team and not individually. Our main experiment can be summarized as "Given three players with the same skill level and self-reported familiarity with games, will Player A (with personality X) perform better/have more fun when playing with Player B (personality Y) or Player C (personality Z)?" The experiment included self-reported demographics, besides the personality evaluation, while the IPIP questionnaire used a 5-point Likert scale. Players reported their understanding of fun and cooperation, comparing two different games played and choosing one of the levels as being more fun than the other. Given the few participants in the experiment due to the short time frame of the study and the mobility restrictions related to COVID-19 (the experiment took place during Summer 2020), our results come from four different player combinations:

- C1 involved players with similar personalities, apart from a major difference in emotional stability
- C2 and C3 involved players with a minor difference in extroversion
- C4 included players with different ratings in extroversion, emotional stability, and contentiousness

The game played by C1 took the most time to complete, but players were the quickest in finishing the most difficult game puzzle, C2 and C3 performed better as a team and managed to solve all puzzles, while C4 was completed relatively quickly, but players failed to solve the final, most difficult puzzle. Overall, our results showed that Emotional Stability was an important factor when it comes to solving the game puzzles quickly or at all. Actually, major differences across personality factors might not influence the outcome of the game, as long as one of the two players scores well in stability.